\documentclass[12pt,preprint]{aastex}

\newcommand{\beq}{\begin{eqnarray}}
\newcommand{\eeq}{\end{eqnarray}}
\shorttitle{Dark Energy and Strong Gravitational Lenses}
\shortauthors{Djorgovski et al.}

\begin{document}

\title{ Constraining Dark Energy From Splitting Angle Statistic of Strong Gravitational Lenses }
\author{Qing-Jun Zhang, Ling-Mei Cheng, and Yue-Liang Wu}
 \affil{\rm Kavli Institute for Theoretical Physics China, Institute of Theoretical
 Physics, \\
 Chinese Academy of Science, Beijing 100080, P.R. China}
 \email{ylwu@itp.ac.cn}

\begin{abstract}
Utilizing the  CLASS statistical sample, we investigate the
constraint of the splitting angle statistic of strong gravitational
lenses(SGL) on the equation-of-state parameter $w=p/\rho$ of the
dark energy in the flat cold dark matter cosmology. Through the
comoving number density of dark halos described by Press-Schechter
theory, dark energy affects the efficiency with which dark-matter
concentrations produce strong lensing signals. The constraints on
both constant $w$ and time-varying $w(z)=w_0+w_az/(1+z)$ from the
SGL splitting angle statistic are consistently obtained by adopting
a two model combined mechanism of dark halo density profile matched
at the mass scale $M_c$. Our main observations are: (a) the
resulting model parameter $M_c$ is found to be $M_c \sim 1.4$ for
both constant $w$ and time-varying $w(z)$, which is larger than $M_c
\sim 1$ obtained in literatures; (b) the fitting results for the
constant $w$ are found to be $w =-0.89^{+0.49}_{-0.26}$ and $w
=-0.94^{+0.57}_{-0.16}$ for the source redshift distributions of the
Gaussian models $g(z_s)$ and  $g^c(z_s)$ respectively, which are
consistent with the $\Lambda \rm CDM$ at 95\% C.L; (c) the
time-varying $w(z)$ is found to be for $\sigma_8 = 0.74$: $(M_c;
w_0, w_a)=(1.36; -0.92, -1.31)$ and $(M_c; w_0, w_a)=(1.38; -0.89,
-1.21)$ for $g(z_s)$ and $g^c(z_s)$ respectively, the influence of
$\sigma_8$ is investigated and found to be sizable for $\sigma_8 =
0.74\sim 0.90$. After marginalizing the likelihood functions over
the cosmological parameters $(\Omega_M, h, \sigma_8)$ and the model
parameter $M_c$, we find that the data of SGL splitting angle
statistic lead to the best fit results $(w_0,
w_a)=(-0.88^{+0.65}_{-1.03},
 -1.55^{+1.77}_{-1.88})$ and $(w_0, w_a)=(-0.91^{+0.60}_{-1.46},
 -1.60^{+1.60}_{-2.57})$ for $g(z_s)$ and $g^c(z_s)$ respectively.
 \end{abstract}



\section{INTRODUCTION}

Since the direct confirmation of the presence of dark energy by
Type Ia supernovae (SNe Ia) observation (Riess et.al $1998$), the
investigation of its property is one of the most important object
in cosmology. Many theoretical models have been developed to
explain or describe dark energy, which is widely believed as the
main component of the cosmological energy today. By now one key to
the question seems to be the precise measurement to the
equation-of-state parameter $w=p/\rho$ of the dark energy(see a
review paper, Peebles \& Ratra 2002). Compared with the data, the
limit to the value of parameter $w$ is continuously improved by
many experimental groups, including the type Ia supernova(Riess et
al. 2004), the Cosmic Microwave Background(CMB)(spergel et al.
2006), and the weak gravitational lenses(WGL)(Weinberg \&
Kamionkowski 2002).   As a complement to these observations, we
shall utilize the splitting angle statistic of strong
gravitational lenses(SGL) to quantificationally investigate its
constraint on the parameter $w$.

Light lines traversing in the universe are attracted and refracted
by the gravitational force of the galaxies on its path, which
bring us the signal of the SGL effect, one of which is the
multiple images of a single far galaxy. Through comparing the
observed number of lenses with the theoretical expected result as
a function of image separation and cosmological parameters, it
enables us to determine the allowed range of the parameter $w$.
Linder (2004) demonstrated that with the addition of strong
lensing image separation measurements, the estimates for
time-varying $w(z)=w_0+w_az/(1+z)$ from CMB, SNe Ia, and WGL could
be improved modestly. To see that, we shall carefully in this note
investigate the analytic process and the power of SGL data only.
There are numerous works studying on the relation between  DE and
SGL splitting angle statistic (e.g., Porciani \& Madau 2000; Li \&
Ostriker 2002; Kuhlen, Keeton \& Madau 2004), but the exact mass
density profile of dark halos is unknown yet, which produces the
most theoretical uncertainty of the analytic process. The two most
widely-used density profiles are the singular isothermal sphere
(SIS) profile and the Navarro-Frenk-White (NFW) profile. Some
analyses were based solely on the SIS model (e.g., Chae et al.
2002). While from the analyzes by Li \& Ostriker (2002, 2003), and
also by Sarbu, Rusin \& Ma (2001), it gave a convictive
illumination that a combined mechanism of at least two models can
effectively reproduce the observed curve of the lensing
probability $P(>\theta)$ to the image splitting angle $\theta$. To
achieve this, an additional parameter $\rm M_c$ is introduced to
divide the mass scale of dark halos into different parts as a
certain density profile is thought to work only on each part.

The rate of structure growth, which determines the number density
of dark halos as the SGL lenses, is very sensitive to the
normalization parameter of the matter power spectrum, $\sigma_8$.
The value of $\sigma_8$ is often related to the matter density
$\Omega_M$ and constrained by the $\rm CMB$ or cluster abundance
observation. Several years ago, in the $\Lambda \rm CDM$ universe
with $\Omega_M =0.27$,  $\sigma_8$ was found to be larger than 0.9
(e.g., Spergel et al. 2003; Wang \& Steinhardt 1998). Then the
data of Wilkinson Microwave Anisotropy Probe (WMAP) Three Year
(Spergel et al. 2007) provided a noticeably smaller value
$\sigma_8 = 0.74\pm0.05$ in comparison with the first-year data
$\sigma_8 = 0.92\pm0.10$. By utilizing the recent clustering
results of XMM-Newton soft (0.5-2 keV) X-ray sources, the X-ray
clustering, and SNIa data, Basilakos and Plionis (2007) showed
that the $\sigma_8\approx0.73$ for $\Omega_M =0.26$ and $w =-
0.90$, which is consistent with the new result. A smaller
$\sigma_8$ implies less structure growth at late times and less
lensing probability for a single source.

In this paper, by using the CLASS statistical sample (Browne et
al.2003), we are going to show how the constraint on the dark energy
equation of state parameters $w(z)$ can reliably be obtained for the
cases with the assumption of constant $w$ and the time-varying
parameterization $w(z) = w_0 + w_a \, z / (1 + z)$, respectively. We
highlight three issues which have not previously been investigated.
First, we investigate the influences of parameter $w$ on the every
step of the lensing process to find the most important point where
the change of parameter $w$ shows its effect. Second, by comparing
the results for different distributions of sources redshift which
have been used in previous works(e.g., Chae et al. 2002; Li \&
Ostriker 2002), we illustrate the quantitative influence that is
introduced by the uncertainty of the source distribution. Third, we
focus on the constraint of the data on the time-varying
parameterization $w(z)$. Our paper is organized as follows: Sect. 2
outlines the cosmological model, the mass fluctuations, and the
Press-Schechter function used in our calculation. Sect. 3 describes
the density profiles and the lensing probabilities. Sect. 4 gives
our data analysis and numerical results. The conclusions are
presented in the last section.

\section{BASIC CONSIDERATIONS}

In this section we shall describe some ingredients used in our
calculations.

\subsection{Cosmological Model and Mass Fluctuations}

 During the last two decades more and more
observational evidence suggests that our universe at present is
accelerated expanding and  dominated by a spatially smooth
component with negative pressure, so called dark energy. Besides
the common cosmological constant supposition, an attractive
alternative candidate for dark energy is the potential energy of a
slow-varying scalar field, which is conveniently parameterized
through $w = p/\rho$. The conventional scalar field models, i.e.
quintessence models,  with a positive kinetic energy term in the
field Lagrangian require $w \geq -1$ (e.g., Ratra \& Peebles 1988;
Caldwell, Dav{\'e}, \& Steinhardt 1998), and phantom dark energy ,
adopting alternatively a negative kinetic energy term, gives the
parameter space $w < -1$ (e.g., Caldwell 2002; Carroll, Hoffmann,
\& Trodden 2003; Cline, Jeon, \& Moore 2003 ). To fit the data of
the observation and give the allowed parameter space, we use two
typical parameterizations, i.e., constant $w$ and time-varying
$w(z)$, as follows:
\begin{eqnarray*}
& & {\rm Case I }:\ \  w = constant \\
& & {\rm Case II}:\  w(z) = w_0 + w_a \, z / (1 + z), \quad w_a =
d w(z) / d z |_{z=0}.
\end{eqnarray*}
The simple form $w(z) = w_0 + w_1 \, z$ is not favored because the
value of $w(z)$ runs to infinite when redshift $z$ goes to be
infinite. Throughout this paper, we assume a flat cold dark matter
universe with present matter density relative to critical density
$\Omega_M=0.24$ and the Hubble parameter $h = 0.73$ (Spergel et
al. 2007).

Dark halos of galaxies and galaxy clusters are formed through
linear growth and nonlinear collapse of primordial fluctuations of
matter in the early universe. The time-varying state-of-equation
parameter $w(z)$ is usually realized by models with slow-varying
scalar fields, while these scalar fields of dark energy begin to
cluster gravitationally and contribute to the perturbation
spectrum only at the very large spatial scale $L>100\rm Mpc$,
which corresponds to a very small wavenumber $k<0.01\rm Mpc^{-1} $
(e.g., Ma et al. 1999). In our present considerations, the most
concerned length scale $l<1\rm Mpc$ is far less than the length
scale $L$ and  the concerned fluctuations are obtained by
integrating the mass power spectrum $k^2 P(k)$ up to a larger
value of the wavenumber $k > 1.0 \rm Mpc^{-1}$. Noticing the fact
that the increasing function $k^2 P(k) \propto k^3$ when $k<<1.0
\rm Mpc^{-1}$, so the contributions from the integral region
$k<0.01\rm Mpc^{-1}$ are very small, namely the contribution of
the dark energy cluster to mass fluctuations can be neglected.
Thus we can directly utilize the mass power spectrum of $\Lambda
\rm CDM$ cosmology. In our calculation, the linear CDM power
spectrum is computed by adopting the fitting formulae given by
(Eisenstein \& Hu 1999)
\begin{eqnarray}
   \Delta_k(k,z)\equiv  {k^3\over 2\pi^2} P(k,z) = \delta_H^2 \left({ck\over
         H_0}\right)^{3+n} T^2(k)\, {{\cal D}^2(z)}\,,
     \label{cdm_pow}
\end{eqnarray}
The initial power spectrum index $n$ is fixed to be $n \equiv 1$.
$T$ is the transfer function
\begin{eqnarray}
     T &=& {L\over L + C q_{eff}^2}\,,
\end{eqnarray}
with
\begin{eqnarray}
     L &\equiv& \ln \left(e + 1.84 q_{eff}\right)\,, \hspace{1cm}
           q_{eff}\equiv {k\over \Omega_M h^2\, {\rm Mpc}^{-1}}\,,\\
     C &\equiv& 14.4 + {325\over 1+ 60.5 q_{eff}^{1.11}}\,.
\end{eqnarray}

The parameter $\delta_H$ is the amplitude of perturbations on the
horizon scale today and related to the rms density fluctuations in
spheres of radius $r_8 = 8 h^{-1} {\rm Mpc}$, so called $\sigma_8$
by:
\begin{eqnarray}
     \delta_H = \frac{\sigma_8}{ \left[\int_0^{\infty} {dk\over k} \Delta_k(k,0) W^2(kr_8)\right]^{1/2} }
     = {\sigma_8 \over \left[\int_0^{\infty} {dk\over k}
            \left({ck\over H_0}\right)^{3+n} T^2 \left(k\right) W^2
            \left(k r_8\right)\right]^{1/2}}
\end{eqnarray}
where $W(kr)$ is the top-hat window function:
$W(kr)=3\left[{sin(kr)\over(kr)^3} -{cos(kr)\over(kr)^2} \right].$
To show the power of SGL data only, we don't use any analytic
fitted form of the parameter $\sigma_8$, expressed by $\Omega_M$
and $w$. Alternatively , unless special clarification we choose to
normalize the power spectrum to $\sigma_8 = 0.74$, the best-fit
value given by WMAP Three Year data (spergel et al. 2006).

The linear growth function ${\cal D}(z)$ is  proportional to the
linear density perturbation $\delta = \delta \rho_M/\rho_M$. The
evolution of linear perturbation is:
 \beq
 \ddot{\delta} + 2{\dot{a}\over a} \dot{\delta} = 4 \pi G \rho_M
\delta
 \eeq
where $a$ is the scale factor $a=(1+z)^{-1}$, dot means derivative
with respect to physical time $t$, the background matter density
$\rho_M = \rho_0 (1+z)^3$, $\rho_0=\Omega_M \rho_{\rm crit,0}$ and
$\rho_{{\rm crit},0}=3 H_0^2/(8\pi G)$ is today's critical mass
density in the universe. Then with the definition $D(z) \equiv
\delta(z)/\delta(z=0)$, we can obtain the equation of ${\cal
D}(a)$:
 \beq
 {d^2 {\cal D} \over da^2} = {3 \over 2}{\Omega \over a^2} {\cal
 D} - {3 \over 2 a}{d {\cal D} \over da}[1-w(a)(1-\Omega)]
  \label{df}
  \eeq
where $\Omega$ is the matter density parameter $\Omega = \Omega_M
(1+z)^3 / (H/H_0)^2$ and
 \[ {H \over H_0} = {\dot{a} \over a H_0} = \sqrt{\Omega_M(1+z)^3+ \Omega_{DE} \,{\rm
 Exp}\left( \int^z_0 dz' \,3[1+w(z')]/(1+z')\right) }\]
Here $H_0$ is the present Hubble constant $H_0=100h \,\rm km
\,s^{-1} Mpc^{-1}$ and $\Omega_{DE}=(1-\Omega_M)$ for a flat
universe. With two boundary conditions: ${\cal D}(a)|_{a=0}=0$ and
${\cal D}(a)|_{a=1}=1$, Equation (\ref{df}) can be calculated out
 numerically. The second boundary condition means ${\cal D}(a)$ is normalized
to ${\cal D}(a=1)=1$.
\subsection{Mass Function and Spherical Collapse Approximation}

 According to the Press-Schechter theory, the comoving number
density of dark halos virialized by redshift $z$ with mass in the
range $(M,M+dM)$ is given by
\begin{eqnarray}
     n(M,z)\,dM = {\rho_0\over M}\, f(M,z)\, dM\,. \label{p-s0}
\end{eqnarray}
 $f(M,z)$ is the Press-Schechter function. We utilize the modified
 form by Sheth $\&$ Tormen (1999)
 \begin{eqnarray}
     & & f(M,z) = - {0.383\over \sqrt{\pi}}  {\delta_c\over \Delta^2} {d\Delta\over dM}
 \left[1+\left({\Delta^2\over0.707 \delta_c^2 }\right)^{0.3} \right] \times {\rm
 exp} \left[- {0.707\over2} \left({\delta_c\over
 \Delta}\right)^2\right]\,,
     \label{p-s} \\
 & & \Delta^2 (M,z) = \int_0^{\infty} {dk\over k} \Delta_k(k,z) W^2(kr)
\end{eqnarray}
where $\Delta$ is the variance of the fluctuations in a sphere
containing a mean mass $M$, and $M$ is related to the length scale
$r$ via
\[ M= \frac{4\pi}{3} r^3 \rho_{M}. \]
The parameter $\delta_c (z)$ is the linear overdensity threshold
for a spherical collapse by redshift $z$. The matter with the
overdensity in a certain scale of the early universe would undergo
the density growth and the spatial scale reducing. When its
average matter density reaches $\delta_c (z)$, virialization
starts and  then a dark halo is formed. In this paper we follow
Wang \& Steinhardt (1998) and Weinberg \& Kamionkowski (2003) to
calculate the $\delta_c (z)$. Under the approximation of spherical
tophat collapse and labeling $R$ as the spatial length scale for a
halo with a certain mass, the collapse process is determined by
the Friedmann equation
 \beq
\left(\frac{\dot{a}}{a}\right)^2=\frac{8\pi G}{3}(\rho_M +
\rho_{DE})\, , \label{stoph1}
 \eeq
and the time-time component of the Einstein equations
 \beq
\frac{\ddot{r}}{r} = -4\pi G\left[\left(w+\frac13\right)\rho_{DE}
+ \frac13\rho_{halo}\right] ,  \label{stoph2}
 \eeq
where  $\rho_{DE}$ is the energy density of dark energy and
$\rho_{halo}$ is the uniform matter density in the scale $r$ .
When the scale factor $a(z)$ is very small, i.e., $a(z) \to 0$ or
$z \sim z_0 \to \infty $, the equivalent linear overdensity
threshold $\delta_c(z)$ at $z_0$ can approximately be evaluated
through the dark halo density $\rho_{halo}$ and the background
matter density $\rho_M$, i.e., $\delta_c(z_0) \simeq (\rho_{halo}
/ \rho_M -1)$. Then utilizing the two boundary conditions:
$r(a)|_{a=0}=0$ and $dr/da|_{a=a_{ta}} = 0$ with $a_{ta}$ the
scale factor at the turn-around time, the function $\delta_c (z)$
can be calculated out numerically as follows
\[
\delta_c (z) =\delta_c(z_0) D(z)/D(z_0),\quad \delta_c(z_0)\simeq
\left(\rho_{halo}/ \rho_M -1 \right)
\]
Note that the turn-around time $t_{ta}$ is determined through the
virial time $t_{vir}$ when the overdensity matter starts to form
dark halos, i.e., $t_{ta} = t_{vir}/2$, which is corresponding to
virial redshift $z$ according to the Press-Schechter theory. In
this sense, the boundary condition  $dr/da|_{a=a_{ta}} = 0$ is
related to the virial redshift $z$, so the resulting scale $r(a)$
of dark halos will depend on the virial redshift $z$. As a
consequence, the dark halo density $\rho_{halo}$ actually relies
on the virial redshift $z$.

Denote $\Delta_{vir} \equiv \rho_{halo}/\rho_M$ the ratio of the
cluster to the background density. Then the nonlinear overdensity
$\Delta_{vir}(z)$ can also be calculated out directly from the
Eq.(\ref{stoph1}) and Eq.(\ref{stoph2}) with their two boundary
conditions and the $\delta_c(z_0)$ given above.

 In Fig. 1, we plot the Press-Schechter function
$f$ against the mass $M (10^{15} h^{-1} M_{\odot})$ of dark halos
at redshift $z = 0.0$, $1.5$, and $3.0$, respectively. The dash,
solid, dot, and dash dot curves are for the typical cases of  $w =
-0.5$, $w = -1.0$, $w = -1.5$, and $w(z) = -1.0 - z/(1+z)$. Fig. 2
shows the P-S function $f$ as a function of the redshift $z$ with
the dark halo mass $M(10^{15}h^{-1} M_{\odot})$ $=0.001$, $0.1$,
and $10$ , respectively. Notice the differences of the ordinate
length scales. From the two figures, we can see  that for larger
$M$ and smaller $z$, the P-S function $f$ is more sensitive to the
change of the parameter $w$. The explanation is clear: (1) more
time passed for the fluctuations to evolve and form dark halos of
larger mass, so larger $M$ corresponds to later time; (2) in the
early time, the relative proportions of dark energy in the total
energy of universe was small and so is its influence on the
spatial-time geometry.  We can see that even for a small $M \sim
0.001$, the change of $w$ can bring somewhat significant shift of
function $f$ in the most range of redshift $z$. This produces the
most power of SGL data to constrain the parameter $w$.

\section{DENSITY PROFILE AND LENSING PROBABILITY}

  The SGL lensing efficiency is very sensitive to the density
profile of dark halos: under the same conditions, the efficiency
of SIS model is larger, at least by one order of magnitude, than
that of NFW model for the image separation angle $\delta \theta <
30''$. In this section, we shall discuss the influence of
different $w$ on the lensing probability $P(>\Delta \theta_0)$ for
SIS profile and NFW profile, respectively.

\subsection{SIS Profile as a Lens}

 \hspace{0.55cm}The SIS profile has a simple spherically symmetric form
(Schneider, Ehlers, \& Falco 1992)
 \beq
    \rho(r) = {\sigma_v^2\over 2\pi G}\,{1\over r^2}\,,
    \label{rho}
 \eeq
where $\sigma_v$ is the velocity dispersion, which can be related
to the mass $M$ of a dark halo via $\sigma_{\nu}^2 = GM/2
\;r_{vir}$ after integrating the density function $\rho(r)$ from
$r=0$ to $r=r_{vir}$. Here $r_{vir}$ is the viral radius of a dark
halo, which is commonly defined by demanding that the mean density
within the virial radius of the halo be a factor $\Delta_{vir}$
times larger than the background density, $\rho_M$, i.e. $M = {4
\pi \over 3} \Delta_{vir} \rho_{cr} r_{vir}^3$. Eliminating the
dependance on the $r_{vir}$, we get
 \beq M = {\sigma_v^3\over G}\left({6\over \pi\Delta_{vir} G
          \rho_{cr}}\right)^{1/2}
          \label{msigmav}.
 \eeq

This profile is supported by the observed flat rotation curves of
the spiral galaxies and is widely utilized in the gravitational
lensing anslysis. Due to its symmetry, the lensing analysis is
quite easy. Integrate the density component along the line of
sight and then we get its surface mass density
 \beq
\Sigma(\xi) = {\sigma_v^2\over 2G}\,{1\over\xi}\,,
    \label{sig}
 \eeq
where $\xi \equiv|${\boldmath ${\xi}|$} and {\boldmath ${ \xi}$}
is the position vector in the lens plane. The lensing equation is
given by
\begin{eqnarray}
\mbox{\boldmath $\eta$} = \frac{D^A_S}{D^A_L} \mbox{\boldmath
$\xi$} - D^A_{LS} \mbox{\boldmath $\alpha$} (\mbox{\boldmath
$\xi$}) \;,
\end{eqnarray}
where {\boldmath  ${\eta}$} is the source position. $D^A_S$,
$D^A_L$, and $D^A_{LS}$ are the angular-diameter distances from
the observer to the source, from the observer to the lens and from
the lens to the source. The angle {\boldmath ${
\alpha}$}({\boldmath $\xi$}) is the gravitational deflection
angle. For a circularly-symmetric surface mass density,
$\Sigma(${\boldmath $\xi$})$=\Sigma(\xi)$, images appear on the
plane defined by the observer point, the lens center, and the
source position, and  the angle $\alpha(\xi)$ is given by
\begin{eqnarray}
\alpha(\xi) =  {8 \pi G  \over c^2 \xi  } \,\int^{\xi}_0 \xi'
\Sigma(\xi')\,d\xi'
\end{eqnarray}
To simplify the lensing equation, we define the length scales in
the lens plane and the source plane as
\begin{eqnarray}
    \xi_0 = 4\pi \left({\sigma_v \over c} \right)^2\,
          {D^A_L D^A_{LS}\over D^A_S}\,,
    \hspace{1 cm}
    \eta_0 = \xi_0\,{D^A_S\over D^A_L}\,,
    \label{xi0}
\end{eqnarray}
Then the position vector of a point in the lens plane or source
plane is {\boldmath $\xi$} $= {\bf x} \xi_0$ or {\boldmath $\eta$}
$= {\bf y} \eta_0$. After the reduction, the lensing equation for
an SIS lens is given by
\begin{eqnarray}
    y = x - {|x|\over x}\,.
    \label{lens1}
\end{eqnarray}
It is easy to see that when $|y|\le 1$, i.e. $|x|\le 1$, a single
source has double images with the separation $\Delta x \equiv 2$
and the splitting angle \beq
    \Delta\theta = {\xi_0\over D^A_L} \Delta x = 8\pi\left({\sigma_v \over c} \right)^2
     {D^A_{LS}\over D^A_S} = {8 \pi G \over c^2} {D^A_{LS}\over D^A_S}
     \left({ \pi  M^2 \rho_M \Delta_{vir} \over  6}\right)^{1/3}\,.
    \label{ds1}
 \eeq
Hence the cross section for two images with a splitting angle
$\Delta\theta > \Delta\theta_0$ is given by \beq
    \sigma = \pi \xi_0^2\, \vartheta\left(\Delta\theta -
             \Delta\theta_0\right)
           = \pi \xi_0^2\, \vartheta\left(M-M_0\right)\,,
    \label{sig1}
\eeq where $\vartheta$ is the step function, and $\rm M_0$ related
with $\Delta\theta_0$ can be solved from the equation(\ref{ds1})

The probability for a source at redshift $z_s$ undergoing a
lensing event on account of the galaxies distribution from the
source to the observer can be obtained by dividing the total
lensing cross-section by the area $A(z)$ of the lens plane,
(Schneider et al. 1992) \beq P = \int_0^{z_s}\int_0^{\infty} {d
D_L\over dz} (1+z)^3 n(M,z)\sigma(M,z)\ dM dz\,,
 \label{lp} \eeq
 where $D_L$ is the proper distance from the observer to the lens.

Inserting equation (\ref{p-s0}) and (\ref{sig1}) into equation
(\ref{lp}), we have for the SIS case (Li \& Ostriker 2002)  \beq
    {dP(>\Delta\theta_0) \over dz}  = 16\pi^3 \rho_{{\rm crit},0}\Omega_M
     (1+z)^3\, {d D_L\over dz}\left({D^A_L D^A_{LS}\over
     D^A_S}\right)^2 \nonumber \\
     \times \int^{\infty}_0 {f(M,z )\over M}  \left({\sigma_v \over c}\right)^4
     \vartheta(M-M_0)  dM  .
    \label{dpz2}
 \eeq  Differentiate
this expression with respect to $\Delta
 \theta_0$, we can obtain the
probability density for a source to
 have a double image with splitting angle $\Delta \theta = \Delta
 \theta_0$:
  \beq
    {d^2 P(>\Delta\theta_0)\over d\Delta\theta_0 dz} &=& 16\pi^3 \rho_{{\rm crit},0}\Omega_M
     (1+z)^3\, {d D_L\over dz}\left({D^A_L D^A_{LS}\over
     D^A_S}\right)^2 {f(M_0,z) \over M_0}\left({\sigma_v(M_0) \over c}\right)^4
     {dM_0 \over
    d\Delta\theta_0} \, ,\nonumber \\
{dM_0 \over d\Delta\theta_0} &=& {3 c^2\over16 \pi G} \; \left({
\pi \rho_M \Delta_{vir} \over 6}\right)^{2/ 3} {D_S^A \over
D_{LS}^A} M_0^{1/3}  .
    \label{dpdtSIS}
 \eeq

The proper distance $D^L$ and the angular-diameter distance $D^A$
from the redshift $z_1$ to $z_2$ are calculated via
\begin{eqnarray}
D^L(z_1,z_2) = \int^{z_2}_{z_1} { dz \over (1+z) H(z)} \;, \quad
D^A(z_1, z_2) = {1\over1+z_2}\int^{z_2}_{z_1}{dz \over H(z)}\;.
\end{eqnarray}

\subsection{NFW Profile as a Lens}

 The NFW profile, based on the N-body numerical
simulation of cold dark matter, is a very important approach for
understanding the formation of galaxies and clusters of galaxies
(e.g., Zhao 1996; Hanyu \& Habe 2001).  Its mass density is  given
by (Navarro, Frenk \& White, 1995, 1996, 1997) \beq
    \rho_{\rm NFW}(r) = {\rho_s r_s^3 \over r(r+r_s)^2}\,,
    \label{nfw}
\eeq where $\rho_s$ and $r_s$ are constants:
\begin{eqnarray}
    \rho_s =   {\rho_M \Delta_{vir}\over 3}{c_1^3\over
         f(c_1)} \;,\quad
    r_s = {1\over c_1}\,\left({3 M\over 4 \pi \rho_M \Delta_{vir}
    }\right)^{1/3} \;,
    \label{rs}
\end{eqnarray} with $f(c_1) = {\rm ln}(1+c_1) - c_1/(1+c_1)$. For
the concentration parameter $c_1$, we adopt the fitting formulae
given by Bullock et al. (2001): $c_1 = 9(1+z)^{-1}({\rm
M}/1.5\times10^{13}h^{-1}{\rm M}_{\odot})^{-0.13}$.

Similar to  the case of SIS, we define the position vector in the
lens plane and the source plane as {\boldmath $\xi$} $= {\bf x}
r_s$ and {\boldmath $\eta$} $= {\bf y}\, r_s D^A_S/D^A_L$,
respectively. The surface mass density for the NFW profile is
given by(Li \& Ostriker 2002)
 \beq
  \Sigma(x) = 2 \rho_s r_s
  \int^{\infty}_0 (x^2+z^2)^{-1/2}[(x^2+z^2)^{1/2}+1]^{-2}dz
 \eeq
Then the reduced lensing equation is
\begin{eqnarray}
    y = x - \mu_s {s(x)\over x}\,,
    \label{le1}
\end{eqnarray}
where \begin{eqnarray}
& & \mu_s \equiv 4 \rho_s r_s / \Sigma_{\rm cr},  \nonumber \\
& & \Sigma_{\rm cr} \equiv {1\over 4\pi G}\,{D^A_S\over D^A_L
         D^A_{LS}}, \nonumber \\
& &    s(x) \equiv \int_0^x u du \int_0^{\infty} \left(u^2 +
           z^2\right)^{-1/2} \left[\left(u^2+z^2\right)^{1/2} +
       1\right]^{-2} dz\,,
    \label{gx}
\end{eqnarray} The dimensionless parameter $\mu_s$ determines the size of the
lensing cross-section $\sigma$ for a NFW halo to produce multiple
images: larger $\mu_s$, smaller $\sigma$. The curve of y to x runs
through the coordinate origin and has a extremum point
central-symmetrically on each side, whose coordinates
($x_{cr}$,$y_{cr}$)  are determined by $dy/dx|_{x_{cr}}=0$ and
$y_{cr}=y(x_{cr})$. Thus a single source with a certain $y$ has
multiple images when $|y|\leq y_{\rm cr}$. Once more than two
images are formed, we shall only consider the splitting angle
$\Delta \theta$ between the two outside images. According to Li \&
Ostriker 2002, we shall neglect the variety of $\Delta \theta$
caused only by the movement of y and get $   \Delta x\,(y) \approx
\Delta x\,(y=0) = 2 x_0 $, where $x_0$ is the positive root of
$y(x) = 0$. Then the splitting angle $\Delta\theta$ is given by
\begin{eqnarray}
    \Delta\theta = {r_s\over D^A_L} \Delta x \approx
        {2 x_0 r_s\over D^A_L}\,.
    \label{deth}
\end{eqnarray} and the cross-section for forming multiple images  with
$\Delta\theta>\Delta\theta_0$ is
\begin{eqnarray}
    \sigma\left(>\Delta\theta_0,M,z\right) \approx \pi y_{\rm cr}^2
        r_s^2\,\vartheta\left(\Delta\theta - \Delta\theta_0\right)\,.
    \label{signfw}
\end{eqnarray}

Fig. 3 shows the splitting angle $\Delta \theta$ as the function
of $M(10^{15} h^{-1} M_{\odot})$ in SIS and NFW cases for  $w =
-0.5$, $w = -1.0$, $w = -1.5$, and $w(z) = -1.0 - z/(1+z)$. The
source object is at $z_s = 1.5$, and the lens object is at $z =
0.3$. In Fig. 4 we plot the $\Delta \theta$ against the redshift
of lens $z$ for $M=0.01$, $1.0$, and $100$, respectively. The
source object is at $z_s = 1.5$. The curves of $w(z)$ and $w=-1.0$
almost overlap in both figures.  We can see that the $\Delta
\theta$ produced by a NFW lens is more sensitive to the parameter
$w$ than that of an SIS lens, especially for small $M$. For SIS
case, There are only quite small changes of $\Delta \theta$ for
our different selections of $w$ in both Fig. 3 and Fig. 4.

Fig. 5 gives the cross section $\sigma$ against the redshift of
lens $z$ for $M(10^{15}h^{-1} M_{\odot})=0.01$, $1.0$, and $100$,
respectively.  The source object is at $z_s = 1.5$. For SIS case,
There are not visible changes of $\Delta \theta$ for our different
selections of $w$. For NFW case, the cross section is more
sensitive to the parameter $w$ for smaller $M$ and somewhat
smaller $z$.

Using equations (\ref{p-s0}) , (\ref{lp}), and (\ref{signfw}), we
then get the differential lensing probability for the NFW case (Li
\& Ostriker 2002) \begin{eqnarray}
     {dP(>\Delta\theta_0)\over dz}  &=& \pi \rho_{crit,0}\,\Omega_M (1+z)^3\,
          {dD_L\over dz}\int^{\infty}_0{f\left(M,z\right)\over M}
          y_{\rm cr}^2 r_s^2
          \vartheta(M-M_0) dM\,.
     \label{dpznfw}
\end{eqnarray} Differentiate this expression with respect to $\Delta
 \theta_0$, we can obtain the
probability density for a source to
 have a double image with splitting angle $\Delta \theta = \Delta
 \theta_0$: \beq
    {d^2 P(>\Delta\theta_0)\over d\Delta\theta_0 dz} &=& \pi \rho_{crit,0}\,\Omega_M (1+z)^3\,
          {dD_L\over dz} {f\left(M_0,z\right)\over M_0}
          y_{\rm cr}^2(M_0) r_s^2(M_0)
           {dM_0 \over
    d\Delta\theta_0} \, ,\nonumber \\
{dM_0 \over d\Delta\theta_0} &=& {D_L^A\over2} \left( r_s(M_0)
{dx_0
\over dM_0} + x_0(M_0) r_s(M_0){1/3+0.13 \over M_0}\right)^{-1} , \nonumber \\
{dx_0 \over dM_0} &=& {g(x_0) / x_0 \; \; \;d \mu_s / dM_0 \over 1
- \mu_s(M_0) (g'(x_0)/x_0 - g(x_0)/x_0^2)} \,, \nonumber \\
{d\mu_s \over dM_0} &=& {4 r_s \rho_s \over M_0
\Sigma_{cr}}\left(0.07+ {0.13 c_1^2\over (1+c_1)^2 f(c_1) }\right)
\nonumber \\
g'(x_0) &=& x_0 \int_0^{\infty} \left(x_0^2 +
           z^2\right)^{-1/2} \left[\left(x_0^2+z^2\right)^{1/2} +
       1\right]^{-2} dz\,,
    \label{dpdtNFW}
 \eeq

\section{DATA ANALYSIS AND NUMERICAL RESULTS}

The CLASS statistical sample provided a well-defined statistical
sample with  $N=8958$
 sources. Totally  $N_l=13$
 multiple image gravitational lenses have been discovered and all have image
separations $\Delta\theta<3^{\prime\prime}$ (Browne et al. 2003).
The data informations of the $13$ observed lens systems are not
entire: the source redshift $z_s$ and lens redshift $z$ are both
unknown for $1$ lens systems, only $z_s$ unknown for $4$ lens
systems and only $z$ unknown for $1$ system.

\subsection{Basic Preparations}

The CLASS statistical sample use the flux density ratio $q_r$ of the
multiple lensing images as a selection criterion of a sample:
$q_r=|\mu_+/\mu_-|\leq10$ (Chae et al.2002; Chen 2003a, 2003b),
where the $\mu_+$ and $\mu_-$ are the magnifications of two
(outside) images, respectively. The magnification of a image is
determined by $\mu = \left[ {y \over x}{dy \over dx}\right]^{-1}$.
The parameter $q_r$ reduces the lensing cross section $\sigma$. For
a NFW case, the influence of $q_r$ on $\sigma$ is very small and can
be neglected. For an SIS lens, it needs to  multiply the $\sigma$ by
a factor $(9/11)^2$.

As the exact redshift distribution of the CLASS statistical sample
is unknown, Chae et al.(2002) utilized a Gaussian model with mean
redshift $\langle z_s \rangle = 1.27$ given by Marlow et al. (2000)
to describe the redshift distribution for the unlensed sources of
the CLASS statistical sample. Denote this Gaussian model as
$g^c(z_s)$, its distribution was explicitly plotted in Figure 5 of
Chae (2003), which was obtained from describing the redshift
distribution of the flat-spectrum sources as shown in that Figure 5.
Note that as such an Gaussian model has a physical cut at the point
$z_s = 0$, it is no longer to have the standard form. Unlike the
Gaussian model $g^c(z_s)$ by Chae (2003), here we shall take an
alternative Gaussian model by directly fitting the redshift
distribution of the subsample of CLASS statistical sample given by
Marlow et al. (2000) instead of fitting the redshift distribution of
the flat-spectrum sources in Chae (2003). Taking the general form of
Gaussian model
 \beq
 g(z_s) = {N_s \over \sqrt{2 \pi} \lambda } {\rm exp} \left[- {(z_s-a)^2}\over 2 \lambda^2\right]
 \eeq
with $N_s$ being the normalization parameter $\int^\infty_0
f(x)dx\equiv1$, and requiring the mean value $\int^\infty_0 x
f(x)dx\equiv1.27$ given by Marlow et al. (2000), we then only need
to fit the remaining one parameter. The best fit results are found
to be
\begin{equation}
N_s=1.6125; \; a=0.4224; \; \lambda=1.3761
\end{equation}
For comparison, we also present some results based on two treating
methods appearing in literatures by using the CLASS statistical
sample: the redshift distributions of sources are the average
redshift value $d(z_s) = \delta(z_s - 1.27)$ (see, e.g., Li \&
Ostriker 2001) and $f(z_s) = 0.204 + 0.2979 z_s - 0.1121 z_s^2 +
0.001584 z_s^3 $(see, e.g., Sarbu et al. 2001). Fig. 6 gives curves
of four models as a function of $z_s$ and the histogram of 27 CLASS
subsample from Marlow et al. (2000).

    Before comparing with the CLASS statistical sample, we should consider the
effect of magnification bias $B$, which causes the
overrepresentation of the lensed objects in a flux-limited survey.
The flux distribution of the CLASS statistical sample is
well-described by $N(f)\propto(f/f_0)^\eta$ with $\eta = 2.07 \pm
0.02(1.97 \pm 0.14)$ for $f\geq f_0$($f\leq f_0$) and $f_0=30mJy$
(Chae et al.2002). The analysis process of magnification bias is
determined by the lensing equation and following equations
 \beq
  B &=& {\int_{A_m}^\infty {dA\over A} \; p(A)\; \int_{f_0}^\infty
   N(f/A)df \over \int_{f_0}^\infty df N(f) } \nonumber \\
 A &=& \Sigma A_i \nonumber \\
   &=& \Sigma {x_i dx \over y  dy} = \Sigma {x_i/y  \over
   dy/dx|_i} \\
 p(A) &=&{ p(y) \over |dA/dy|} ; \,\,\, p(y) = \alpha y;
 \,\,\,\int^{y_{max}}_0 p(y)dy = 1,
 \eeq where $x$ and $y$ are the position parameters on the lens
 plane, $A$ is the total amplification of the multiply-imaged
 sources, $A_m$ is the minimum value of $A$ and $p(A)$ is the
 probability density for the amplification $A$, $y_{max}$ is the maximum
 value of a source' position on the lens plane.
 $\alpha$ is the normalization constant of probability function $p(y)$:
 for the SIS case $y_{max}=1$, $\alpha = 2$ and for the NFW case
 , it can be calculated out numerically by using $dy/dx|_{y_{max}}=0$. Then for the SIS case,
 after considering the influence of flux density
ratio $q_r$, one needs only multiply the lensing probability by a
constant factor $3.36$. For the NFW case, B can be obtained from
the numerical calculation. $P_{\rm obs}(>\Delta\theta )$ and
$dP_{\rm obs}(>\Delta\theta )/d \Delta \theta $ are related to $P$
by an integration
\begin{eqnarray}
   p(w) \equiv P_{\rm obs}(>\Delta\theta ) = \int\int B\, {d P(>\Delta\theta )\over
       dz}\, \varphi(z_s) dz dz_s\,,
    \label{pobs}
\end{eqnarray}
and
\begin{eqnarray}
   q(w) \equiv {dP_{\rm obs}(>\Delta\theta ) \over d\Delta \theta } = \int\int B\, {d^2 P(>\Delta\theta )\over
       d\Delta\theta  dz}\, \varphi(z_s)  dz dz_s\,,
    \label{pobs2}
\end{eqnarray}
 with $\varphi(z_s)$ is the redshift
distribution of sources.

We shall compare the theoretical results of the SIS case and NFW
case with the CLASS statistical sample. Fig. 7 shows the lensing
probability $P(>\Delta \theta)$ as a function of the splitting
angle $\Delta \theta$ for the source redshift distribution
$d(z_s)$, $f(z_s)$, $g^c(z_s)$ and $g(z_s)$. The thickest line in
each panel is induced from the $13$ observed lensing data. It is
seen that the lensing probability $P(>\Delta \theta)$ of SIS case
and NFW case are both sensitive to the parameter $w$, especially
for NFW case. When parameter $w$ increases, the values of
$P(>\Delta \theta)$ for both SIS and NFW cases clearly increase in
the whole concerned ragion of $\Delta \theta$. Thus it is feasible
to constrain the parameter $w$ from the SGL splitting angle data.
The figure shows that  the SIS model can only reproduce the data
curve at small $\Delta \theta < 1.5''$. When consider the rapid
decline of $P(>\Delta \theta)$ from the data line at large $\Delta
\theta$, a combined mechanism of SIS and NFW model is needed to
explain the whole experimental curve. Define a new model parameter
$M_c$ as Li \& Ostriker (2001): lenses with mass $M<M_c$ have the
SIS profile, while lenses with mass $M > M_c$ have the NFW
profile. Then the differential probability
 \[ dP/dM =
dP_{SIS}/dM\, \vartheta(M_c - M) + dP_{NFW}/dM\,\vartheta(M -
M_c)\] where $\vartheta$ is the step function, $\vartheta(x-y)=1$,
if $x>y$ and 0 otherwise. Because the splitting angle $\Delta
\theta$ is directly proportional to the mass $M$ of lens halos,
the contribution to large $\Delta \theta$ of SIS profile is
depressed by $M_c$. The lens data require a mass threshold $M_c
\sim 10^{13}h^{-1}M_{\odot}$, which is consistent with the halo
mass whose cooling time equals the age of the universe today. In
this note we shall use such a two-model combined mechanism to
calculate lensing probabilities.

In Fig. 7,  One can also find the influences of different source
distributions on the lensing probability $P(>\Delta \theta)$: the
patterns of the function curves are hardly changed, but the
function values for the same $\Delta \theta$ slightly increase
from $d(z_s)$ to $f(z_s)$.

\subsection{Constraint on $w$}

 We now come to our main purpose i.e., utilizing the SGL splitting angle
statistic data from strong gravitational lenses to constrain the
equation-of-state parameter $w$ of dark energy. For that, we define
the likelihood function as \beq {\rm L}(w) =
(1-p(w))^{N-N_l}\prod_{i=1}^{N_l} q_i(w). \label{likh}\eeq $p(w)$
represents the model-predicted lensing probabilities $P(>0.3'')$ of
a source with the redshift distribution $\varphi(z_s)$ and can be
calculated by using equations (\ref{dpz2}), (\ref{dpznfw}), and
(\ref{pobs}). Here $\Delta \theta \geq 0.3''$ is an observational
selection criteria. $q_i(w)$ is the model-predicted differential
lensing probabilities $dP(> \Delta \theta_i )/d \Delta \theta_i$ of
$i^{th}$ observed lens system with the splitting angle $\Delta
\theta_i$ and can be calculated by using equations (\ref{dpdtSIS}),
(\ref{dpdtNFW}), and  (\ref{pobs2}). To utilize the data
informations adequately, we multiply the equation (\ref{pobs}) by
$\delta(z-z^i)$ or/and $\delta(z_s-z_s^i)$(replace $\varphi(z_s)$)
for the $i^{th}$ observed lens system whose lens redshift $z^i$
or/and source redshift $z_s^i$ is/are known. For the unknown $z^i$
or $z_s^i$, we just integrate it out. Compared with utilizing the
curve of lensing probability $P(>\Delta \theta)$ as a function of
the splitting angle $\Delta \theta$ in the last subsection, which is
introduced from lensed signals only, by using the likelihood
function equation (\ref{likh}), the unlensed signals are utilized
and their influences are quite important due to the large
exponential number $(N-N_L)=8945$.

Firstly we discuss the possible constraints on the model parameter
$M_c$ and the constant $w$. Under the given cosmological
parameters $(\Omega_M, h, \sigma_8) = (0.24, 0.73, 0.74)$, Fig. 8
shows the 68\% C.L. and 95\% C.L. allowed regions  from the CLASS
statistical sample for the source redshift distribution $g(z_s)$,
$g^c(z_s)$, $d(z_s)$ and $f(z_s)$, respectively. The crosshairs in
three panels mark the best-fit points $(w, M_c)=(-0.89, 1.37)$,
$(-0.94, 1.36)$, $(-1.4, 1.68)$ and $(-0.73, 1.27)$ from left to
right, where the unit of $M_c$ is $10^{13} h^{-1}M_{\odot}$. The
source redshift distributions $g(z_s)$ and $g^c(z_s)$ give nice
constraints, while $d(z_s)$, which is not proper redshift
distributions, provides somewhat strange unexpected results. The
redshift distribution $f(z_s)$, which has a larger part of
galaxies at high redshift, prefers a larger $w$ and smaller $M_c$.
From now on we will only consider the $g(z_s)$ case and $g^c(z_s)$
case, which are extracted from the subsample of the CLASS
statistical sample (Marlow et al. 2000). Our best fit result of
$M_c \approx 1.40$ for both $g(z_s)$ case and $g^c(z_s)$ case is
larger than the value $M_c\approx1.0$ obtained by Li \& Ostriker
2002. The $95\%$ C.L. allowed regions of parameter $w$ for
$g(z_s)$ case and $g^c(z_s)$ case are from $-0.18$ to $-1.45$ and
from $-0.11$ to $-1.85$, which are consistent with the $\Lambda
\rm CDM$ cosmology. The Fig.2 in Chae (2007) shows a much negative
result of the parameter $w$. Our consideration here is based on
the specially selected cosmological parameters, and also we have
used the two-model combined mechanism, namely the utilization of
model parameter $M_c$. As a consequence, our results avoid the
large absolute value of parameter $w$.

In Fig. 9, we show the constraint for the parameters $(w_0, w_a)$
appearing in a time-varying equation-of-state $w(z)=w_0+w_a
z/(1+z)$ under the given cosmological parameters $(\Omega_M, h) =
(0.24, 0.73)$ for both $g(z_s)$ and $g^c(z_s)$ cases, and with two
different values of $\sigma_8 = 0.74$ and $\sigma_8 =0.90$. The
crosshairs mark the best-fit points $(M_c; w_0, w_a)=(1.36; -0.92,
-1.31)$ for $g(z_s)$ case and $(M_c; w_0, w_a)=(1.38; -0.89,
-1.21)$ for $g^c(z_s)$ case when $\sigma_8 = 0.74$,  and $(M_c;
w_0, w_a)=(1.56; -0.81, -2.5)$ for $g(z_s)$ case and $(M_c; w_0,
w_a)=(1.54; -0.83, -2.22)$ for $g^c(z_s)$ case when $\sigma_8 =
0.90$. The parameter $\sigma_8$ has significant influences on the
mass power spectrum and the number density of dark halos, and our
results show that the best fit $(w_0, w_a)$ are  changed for the
different selections of the parameter $\sigma_8$: when $\sigma_8$
increases from $\sigma_8 = 0.74$ to $\sigma_8 = 0.9$, the best fit
parameters $w_0$ increase moderately and the best fit parameters
$w_a$ have a sizable decrease for both the $g(z_s)$ and the
$g^c(z_s)$ cases.

After marginalizing the cosmological parameters $(\Omega_M, h,
\sigma_8)$  and the critical mass parameters $M_c$ by Monte Carlo
method, we obtain the constraint on $(w_0, w_a)$ in Fig. 10.
 For the three cosmological parameters, we assume the
 Gaussian prior distributions induced from the results of WMAP Three Year Data:
 $(\Omega_M \pm \sigma_m, h \pm \sigma_h, \sigma_8 \pm \sigma_{\sigma_8}) =
 (0.238\pm0.019, 0.73\pm0.03, 0.74\pm0.06)$. For the model
 parameter $M_c$, we integrate it from $1.0$ to $4.0$. The crosshairs mark
 the best-fit point $(w_0, w_a)=(-0.88,
 -1.55)$ for $g(z_s)$ case and $(w_0, w_a)=(-0.91,
 -1.60)$ for $g^c(z_s)$ case. At the $95$\% C.L., our fitting results are
consistent with that of Barger et al. (2006), and the SGL
splitting angle statistic with the source redshift distribution
$g(z_s)$ and $g^c(z_s)$ gives somewhat more negative values for
the parameter $w_a$.

\section{CONCLUSIONS}

From the above analyzes, we have shown how the SGL splitting angle
statistic can be used to quantitatively constrain the
equation-of-state parameter $w$ of dark energy. Though due to the
limited space-time, the difference of the parameter $w$ has few
influences on the splitting angle $\Delta \theta$ and lensing cross
section $\sigma$, while through the comoving number density of dark
halos as sources and lenses described by Press-Schechter theory,
dark energy can affect the efficiency with which dark-matter
concentrations produce strong lensing signals. With a two model
combined mechanism of dark halo density profile, which introduces a
model parameter $M_c$, we have carefully investigated the
constraints on constant $w$ and time-varying $w(z)=w_0+w_az/(1+z)$.
We find the best fit value $M_c \approx 1.4 $ for both $w$ and
$w(z)$, such a value is larger than the value $M_c\approx 1$
obtained by Li \& Ostriker 2002. This is mainly because in our
analyzes both the lensed and unlensed signals have been utilized in
the likelihood function equation (\ref{likh}), while in the analyzes
by Li \& Ostriker 2002, only the lensed signals were be used. The
transition from SIS to NFW characterized by the parameter $M_c$ is
also motivated by the process of baryonic cooling (e.g., Kochanek \&
White 2001), where the parameter $M_c$ was introduced to divide the
cooled(SIS) and uncooled(NFW) halos. The estimated value by Kochanek
\& White (2001) for $h=0.67$ is $M_c \approx 1 \times 10^{13}
M_{sun}$ for the model without a bulge, which is also smaller than
our fitting result. There are several differences between our
calculations and theirs. Firstly, we use a larger Hubble constant $h
= 0.73 > h = 0.67$ and in our Figure 9, we show that for a larger h,
we have a larger fit $M_c$. Secondly, the SIS model used by us has a
larger relative lensing cross section than the exponential disk used
by Kochanek \& White (2001), which gives a larger $M_c$. Thirdly,
the ratio of the cluster to the background density
$\Delta_{vir}=\rho_{halo}/\rho_M > 150$ used by us is much larger
than the value $\Delta \approx 100$ used by Kochanek \& White
(2001), which could also have some influences on the results.
Nevertheless, all the fitting results for $M_c$ are consistent with
appropriate considerations.

With the given cosmological parameters $(\Omega_M, h, \sigma_8) =
(0.24, 0.73, 0.74)$, we have compared the results of constant $w$
corresponding to four kinds of source redshift distributions. It
has been shown that $d(z_s)$ is not suitable for the SGL data
analysis. For the redshift distributions of normalized
Gaussian-type model $g(z_s)$ and Gaussian model $g^c(z_s)$, the
fitting results are $(w, M_c)=(-0.89^{+0.49}_{-0.26},
1.37^{+0.47}_{-0.33})$ and $(w, M_c)=(-0.94^{+0.57}_{-0.16},
1.36^{+0.47}_{-0.34})$ respectively, and the fitting results for
the constant $w$ are consistent with the $\Lambda \rm CDM$ at 95\%
C.L.. For the time-varying $w(z)$, we have firstly investigated
the influence of $\sigma_8$ with the redshift distributions
$g(z_s)$ and $g^c(z_s)$ and found that the fitting results of the
double parameters $(w_0, w_a)$ are changed when $\sigma_8$
increases from $\sigma_8=0.74$ to $\sigma_8=0.9$. With the above
given cosmological parameters, the best fitting results for the
$g(z_s)$ case are $(M_c; w_0, w_a)=(1.36; -0.92, -1.31)$ for
$\sigma_8 = 0.74$ and $(M_c; w_0, w_a)=(1.56; -0.81, -2.5)$ for
$\sigma_8 = 0.9$; and for $g^c(z_s)$ case, the best fit results
are $(M_c; w_0, w_a)=(1.38; -0.89, -1.21)$  for $\sigma_8 = 0.74$
and $(M_c; w_0, w_a)=(1.54; -0.83, -2.22)$ for $\sigma_8 = 0.9$.

 After marginalizing our likelihood functions over the
cosmological parameters $(\Omega_M, h, \sigma_8)$ (by using the
prior probabilities induced from the WMAP Three Year data) and the
model parameter $M_c$, we have obtained a reliable constraint on
the parameters $(w_0, w_a)$. Within the allowed uncertainties, the
results for $w_0$ are consistent with the constraints obtained
from the Type Ia supernovae data (Barger, Guarnaccia \& Marfatia
2006). Our fit results are $(w_0, w_a)=(-0.88^{+0.65}_{-1.03},
-1.55^{+1.77}_{-1.88})$ for $g(z_s)$ case and $(w_0,
w_a)=(-0.91^{+0.60}_{-1.46}, -1.60^{+1.60}_{-2.57})$ for
$g^c(z_s)$ case. It is noticed that the best fitting results based
on the SGL splitting angle statistic favor negative values for the
parameter $w_a$, which differs from the best fitting values
obtained based on the Type Ia supernovae data, where the best
fitting results favor positive values for the parameter $w_a$
(Barger, Guarnaccia \& Marfatia 2006). A combining constraint is
interesting and will be investigated elsewhere.

In conclusion, the quantitative investigation has shown that the
SGL splitting angle statistic can lead to a consistent constraint
on the constant $w$ and the double parameters $(w_0, w_a)$ of the
time-varying dark energy equation of state $w(z)=w_0+w_a z/(1+z)$.
Especially for the allowed range of parameters $(w_0, w_a)$, the
SGL splitting angle statistic does give an interesting bound. It
can be seen from Fig.8 to Fig.10 that the normalized Gaussian-type
source redshift distribution $g(z_s)$ leads to the most stringent
constraints. Though it does not yet allow to obtain a more
accurate constraint, while it can provide a complementarity to
other constraints from Supernovae, cosmic microwave background,
weak lensing.

\section*{Acknowledgments}
 The authors would like to thank R.G. Cai, X.M. Zhang, and Z.H. Zhu for useful
discussions. The authors also acknowledge the useful comments from
referee. This work was supported in part by the National Science
Foundation of China (NSFC) under the grant \# 10821504,10475105,
10491306 and the Project of Knowledge Innovation Program (PKIP) of
Chinese Academy of Science.


\clearpage

\begin{figure}
\epsscale{1.0} \plotone{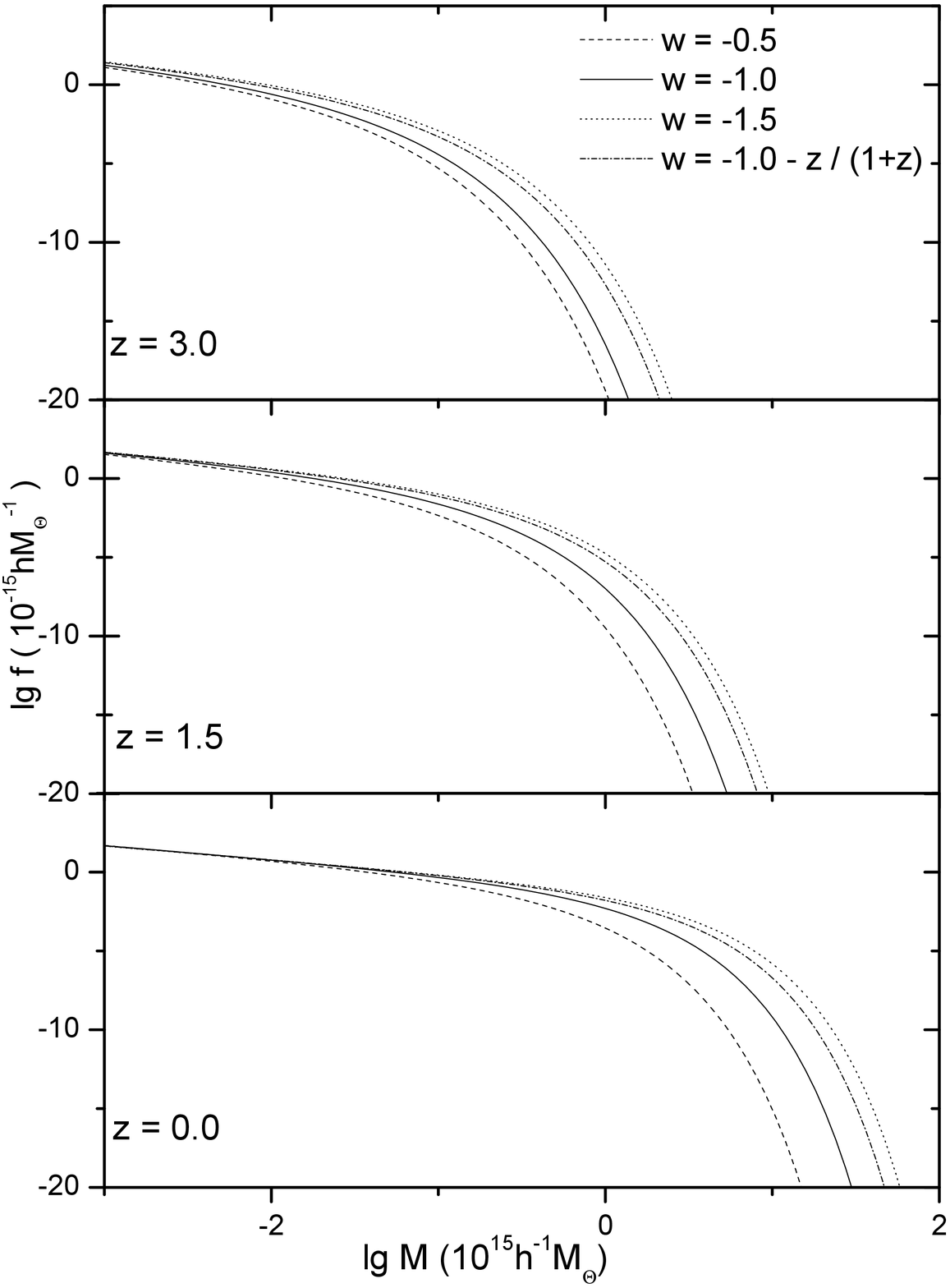}
\figcaption{\small Press-Schechter function $f$ against the mass
$M(10^{15} h^{-1} M_{\odot})$ of dark halos at redshift $z = 0.0$,
$1.5$, and $3.0$, respectively. The dash, solid, dot, and dash dot
curves
 are for the cases of  $w = -0.5$, $w = -1.0$, $w = -1.5$, and $w(z) = -1.0 - z/(1+z)$.}
\end{figure}

\begin{figure}
\epsscale{1.0} \plotone{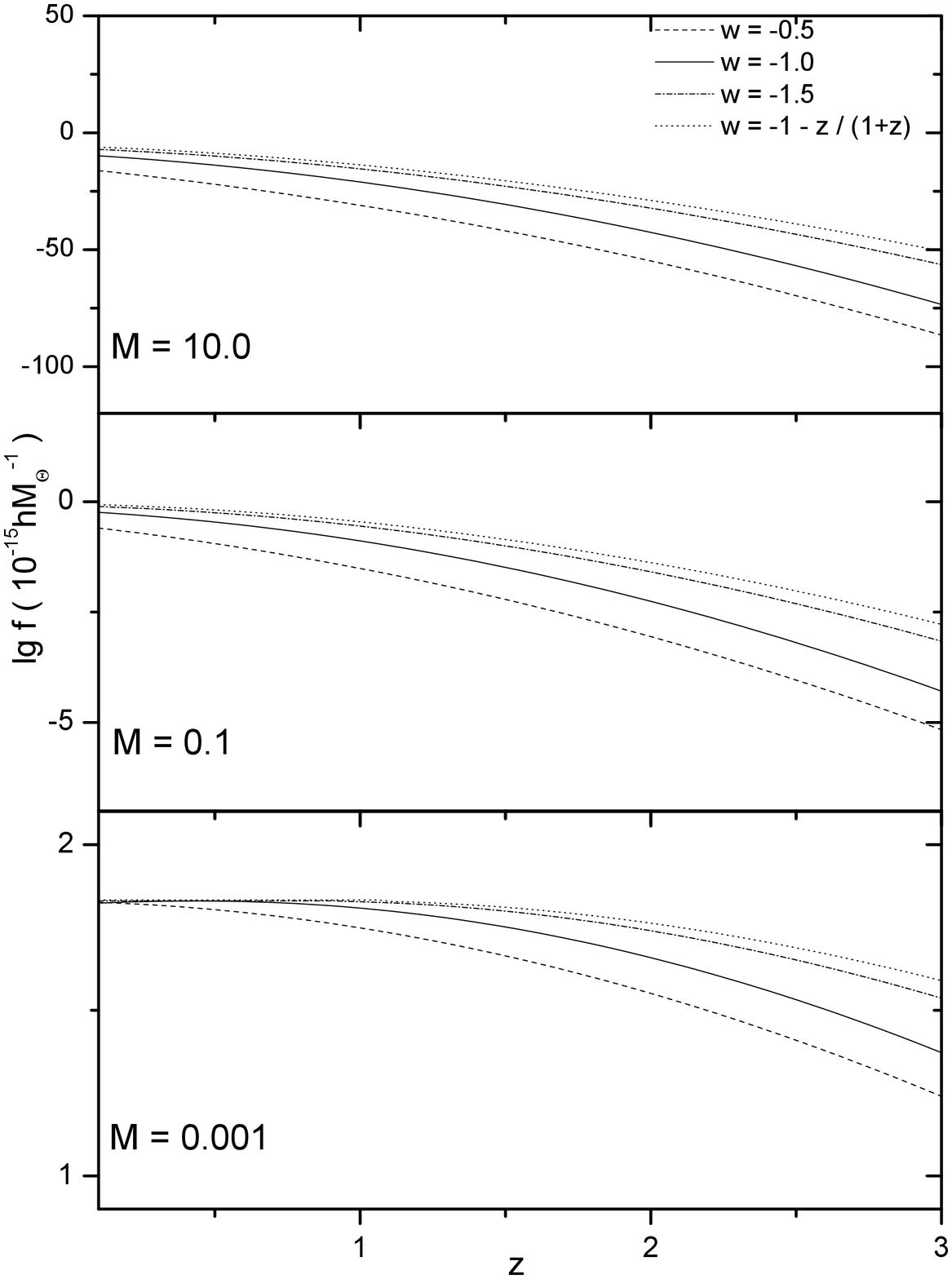}
\figcaption{\small  Press-Schechter function $f$ against the
redshift $z$ for
 different mass $M(10^{15} h^{-1} M_{\odot})$ of dark halos $M=0.01$, $M =1.0$, and $M=100$. The dash(solid, dot or dash dot) curve
 is for the case of  $w = -0.5$ ($w = -1.0$, $w = -1.5$ or $w(z) = -1.0 - z/(1+z))$.}
\end{figure}

\begin{figure}
\epsscale{1.0} \plotone{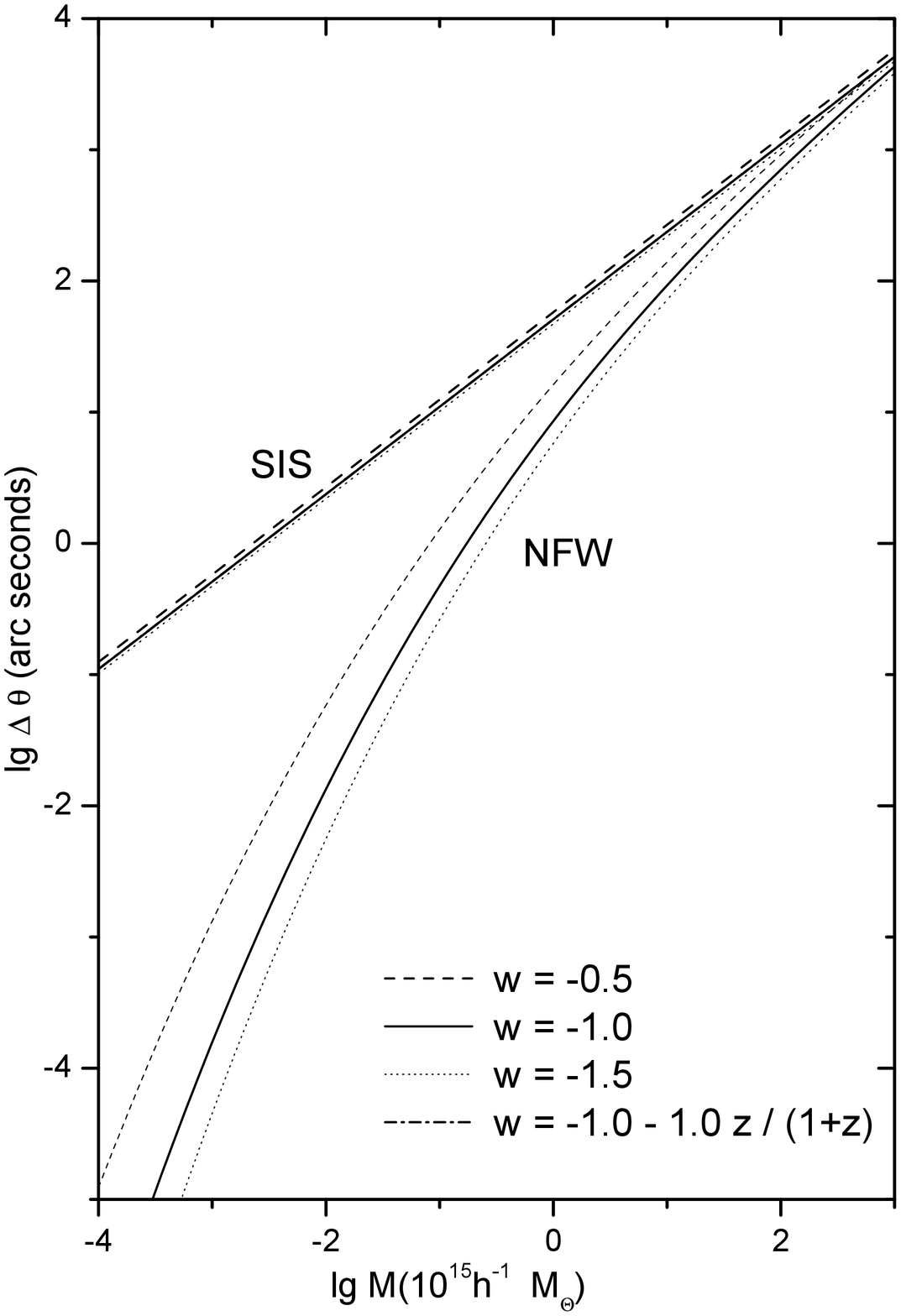}
\figcaption{\small Splitting angle $\Delta \theta$ as the function
of $M(10^{15} h^{-1} M_{\odot})$ in SIS and NFW cases for $w =
-0.5$, $w = -1.0$, $w = -1.5$, and $w(z) = -1.0 - z/(1+z)$. The
source object is at $z_s = 1.5$, and the lens object is at $z =
0.3$. The thick and thin lines show the splitting angle produced
by a NFW lens and an SIS lens, respectively. The dash, solid, dot
or dash dot curve
 is for the case of  $w = -0.5$, $w = -1.0$, $w = -1.5$ or $w(z) = -1.0 - z/(1+z)$. }
\end{figure}

\newpage

\begin{figure}
\epsscale{1.0} \plotone{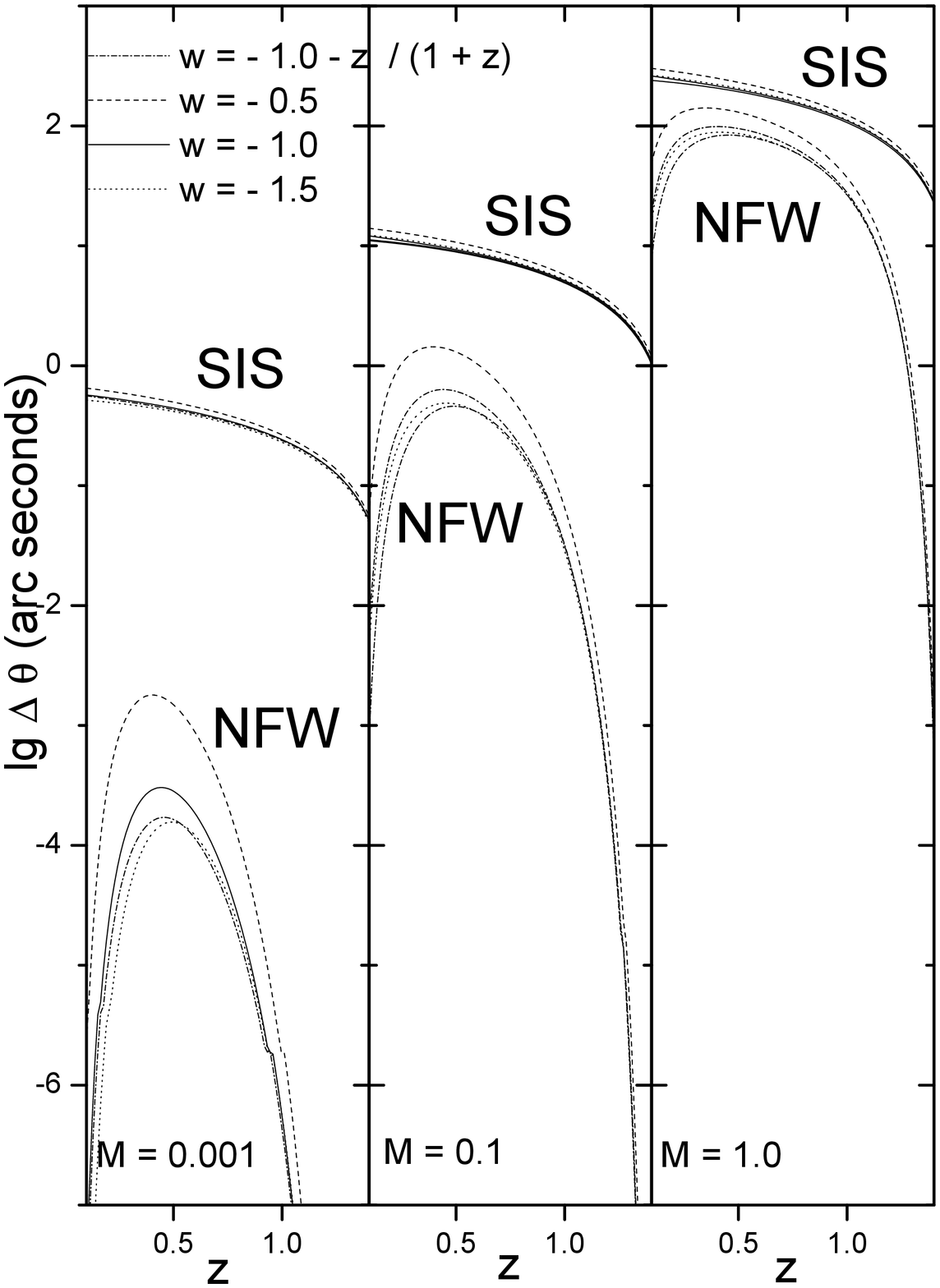}
\figcaption{\small Splitting angle $\Delta \theta$ as the function
of the redshift of lens $z$ for $M=0.01$, $M=1.0$, and $M=100$,
respectively. The source object is at $z_s = 1.5$.  }
\end{figure}

\begin{figure}
\epsscale{1.0} \plotone{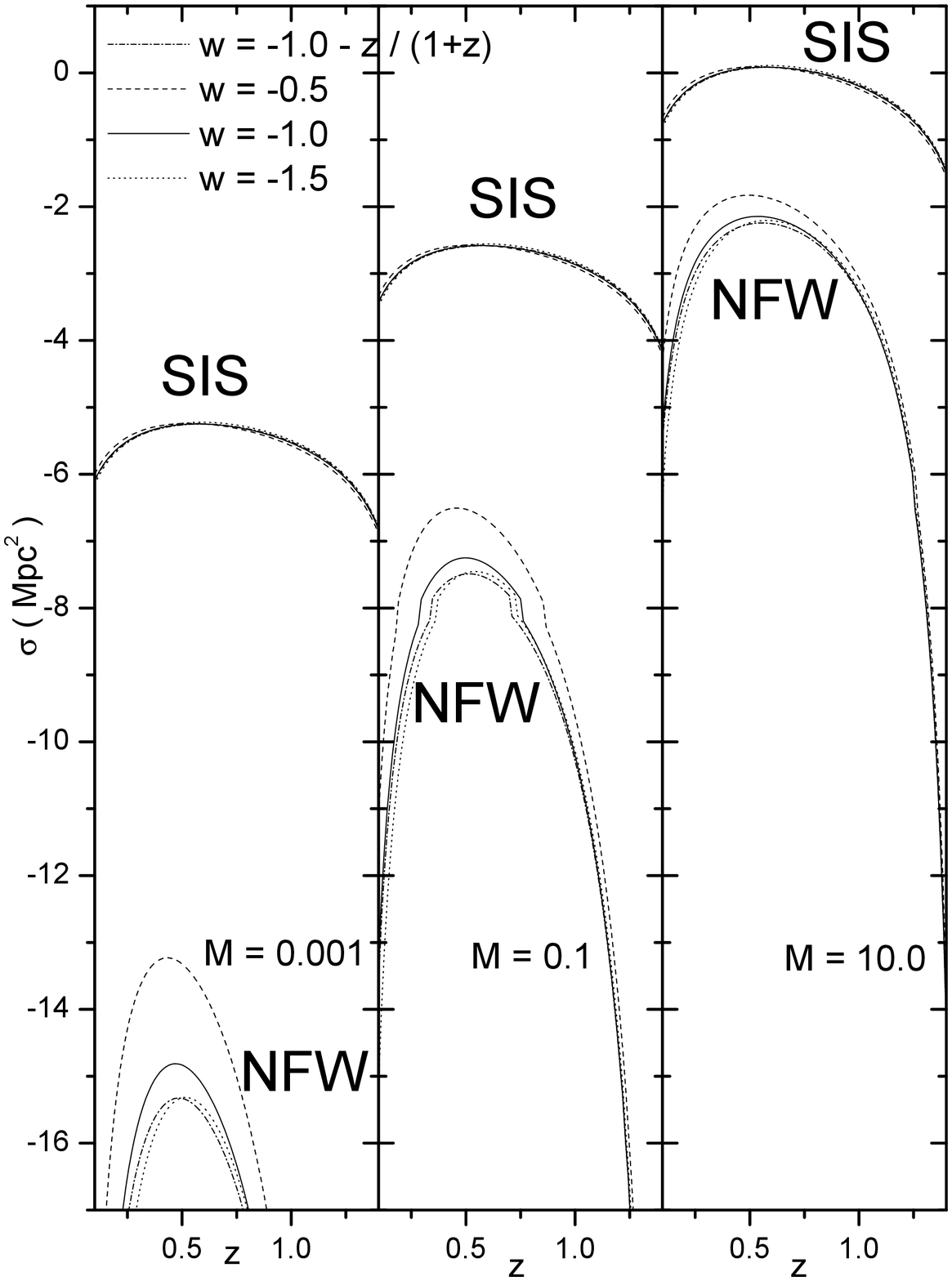}
\figcaption{\small Cross section $\sigma$ against the redshift of
lens $z$ for $M=0.01$, $M=1.0$, and $M=100$, respectively. The
source object is at $z_s = 1.5$.} \end{figure}

\begin{figure}
\epsscale{1.0} \plotone{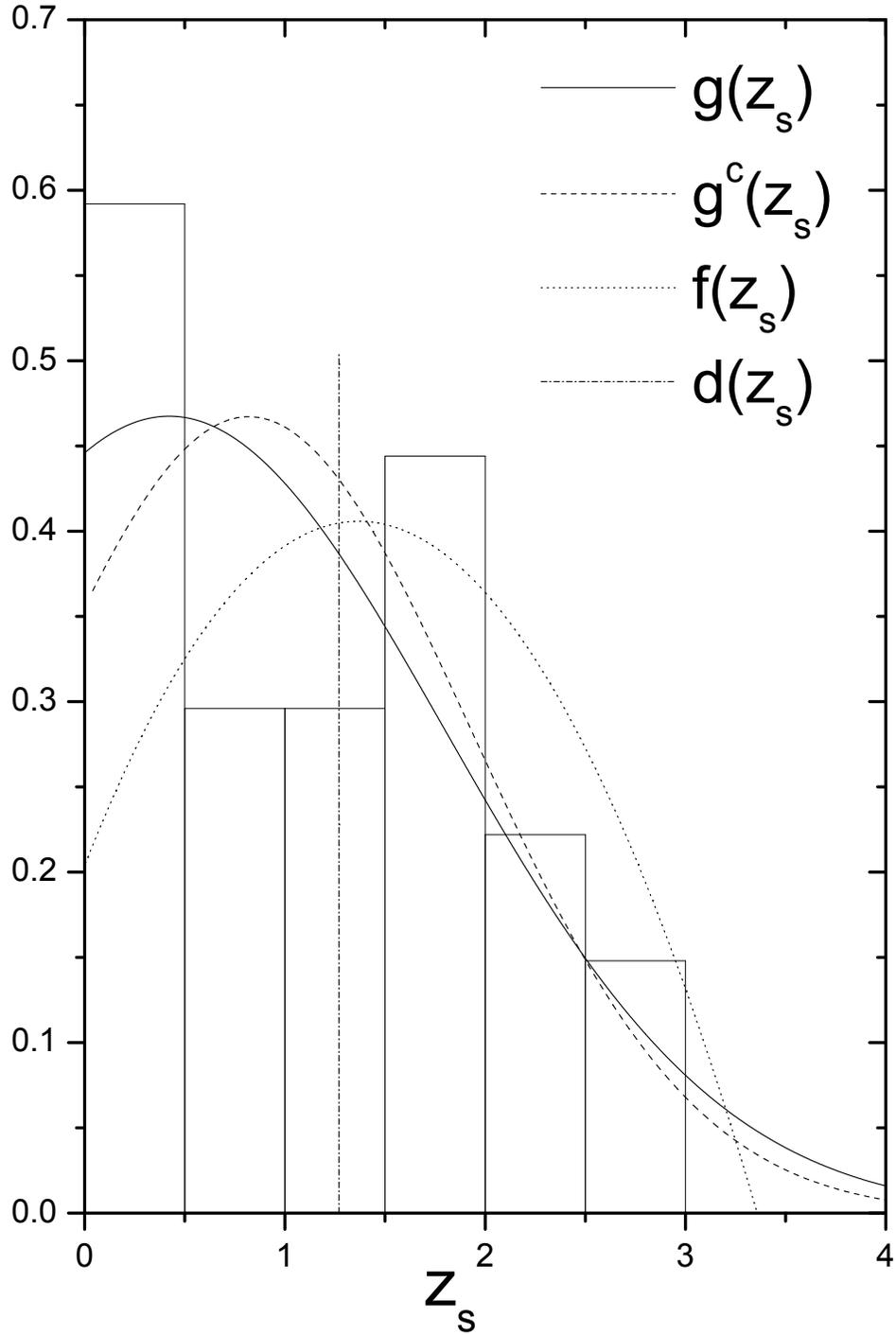}
\figcaption{\small The source redshift distribution functions
against the $z_s$ and the histogram of 27 CLASS subsample from
Marlow et al. (2000). }
\end{figure}


\begin{figure}
\epsscale{1.0} \plotone{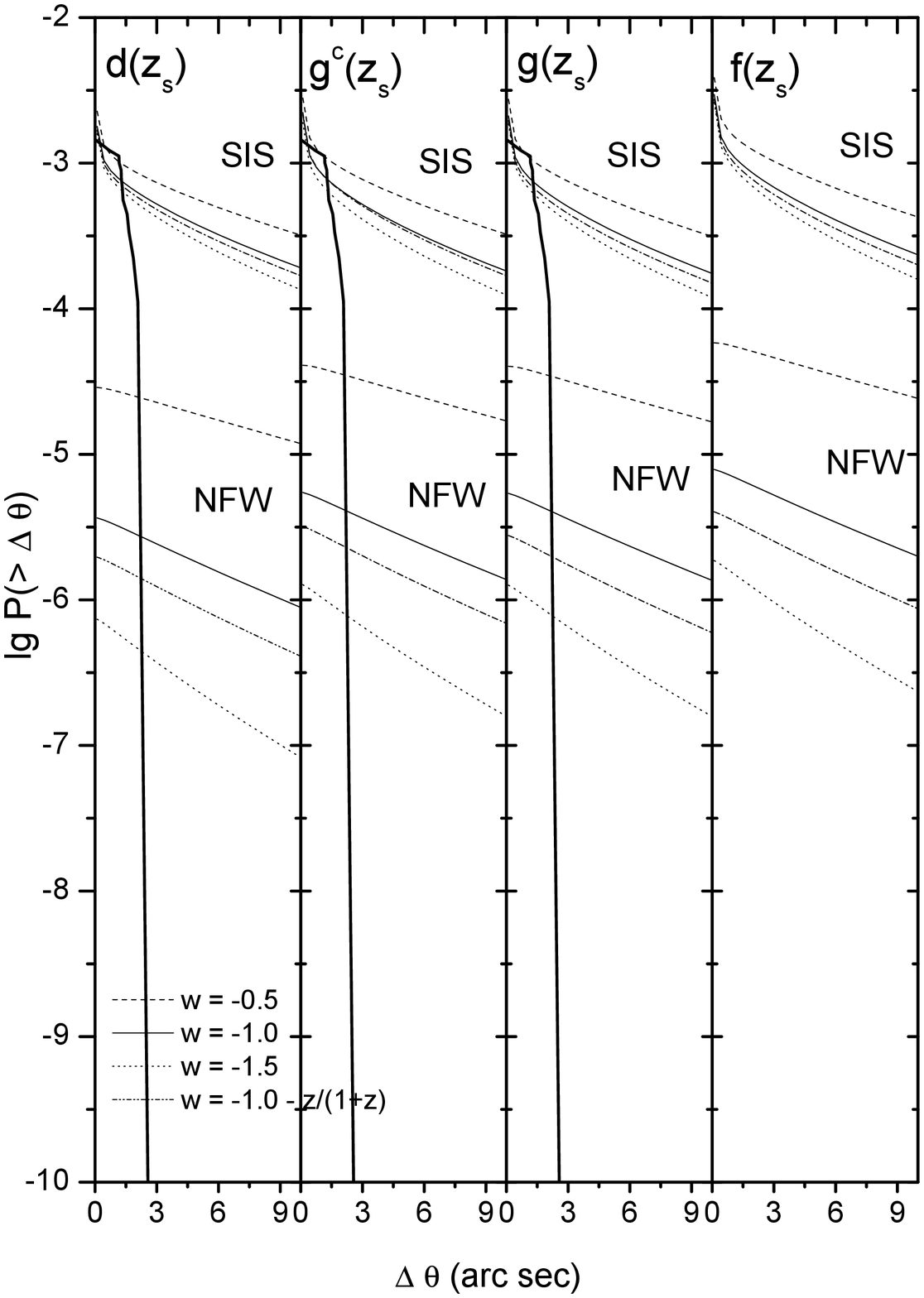}
\figcaption{\small Comparison of the theoretical results of SIS
case and NFW case with the CLASS statistical sample for the four
source redshift distributions, respectively. The CLASS statistical
sample is shown as the thickest line.}
\end{figure}

\begin{figure}
\epsscale{0.9} \plotone{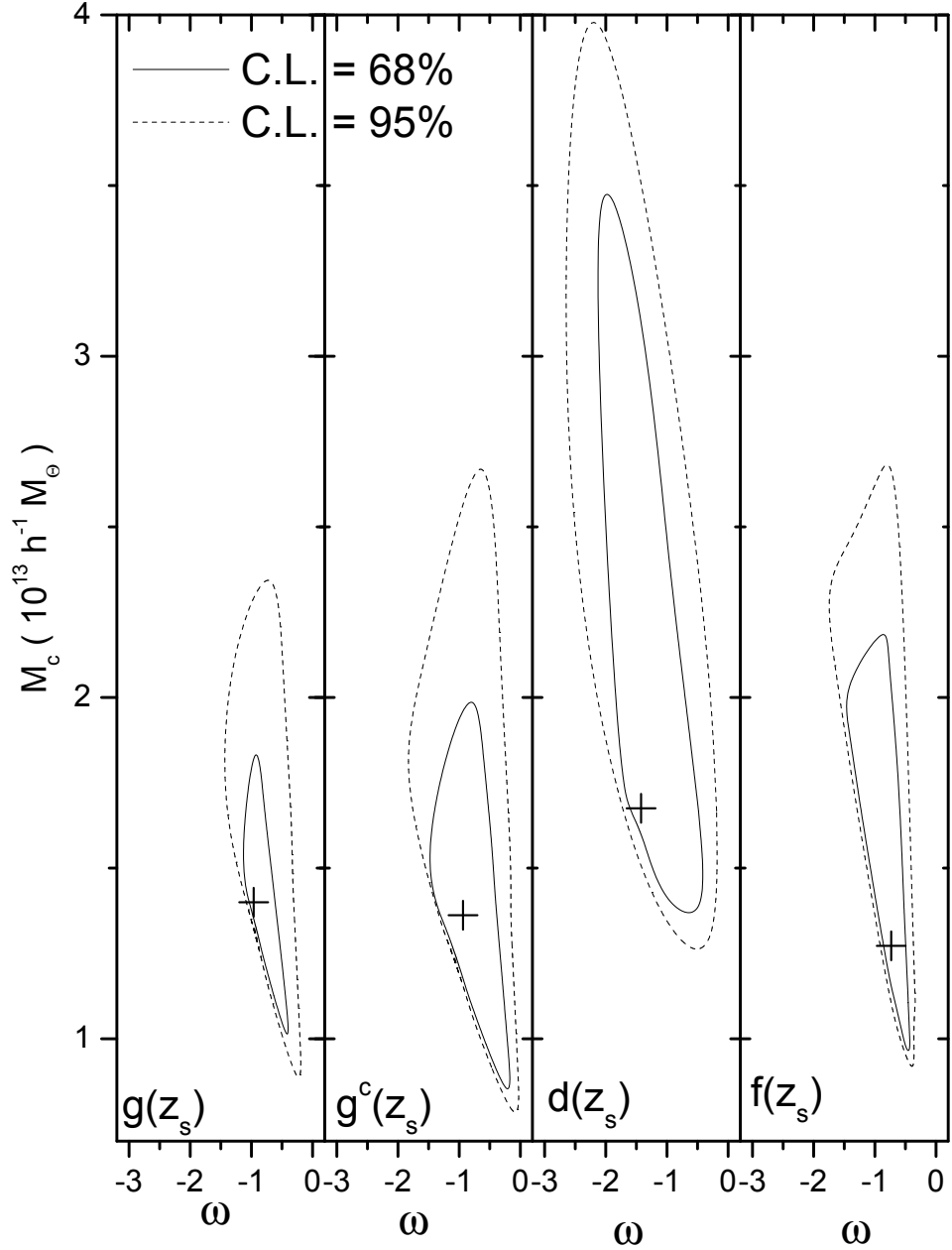}
\figcaption{\small 68\% C.L. and 95\% C.L.  allowed regions from
the CLASS statistical sample for the source redshift distribution
$g(z_s)$, $g^c(z_s)$, $d(z_s)$ and $f(z_s)$, respectively. The
crosshairs in three panels mark the best-fit points $(w,
M_c)=(-0.89, 1.37)$, $(-0.94, 1.36)$, $(-1.4, 1.68)$ and $(-0.73,
1.27)$ from left to right. }
\end{figure}

\begin{figure}
\epsscale{0.8} \plotone{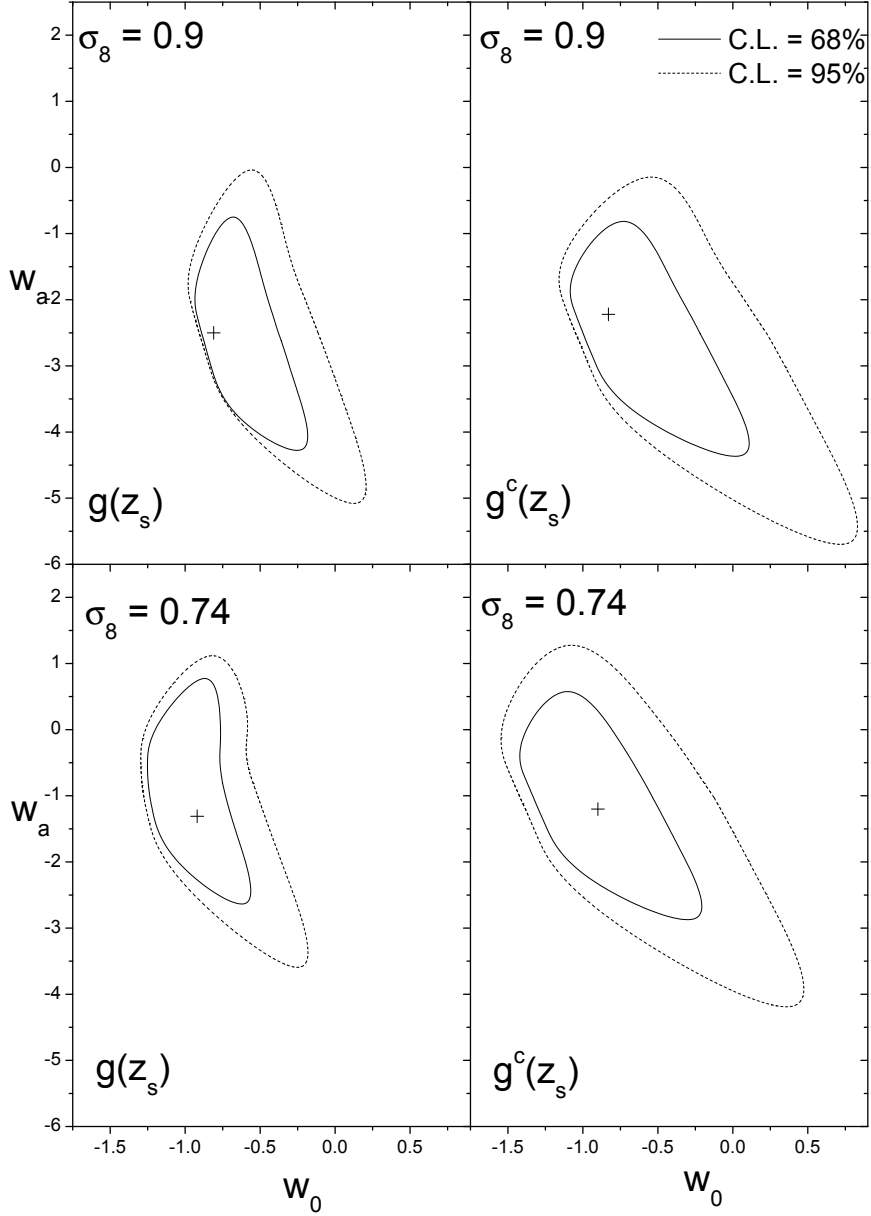}
\figcaption{\small 68\% C.L. and 95\% C.L.  allowed regions from
the CLASS statistical sample for the $f(z_s)$ redshift
distribution with different $\sigma_8 = 0.74$ and $0.9$,
respectively. The crosshairs mark the best-fit points: for the
$g(z_s)$ case are $(M_c; w_0, w_a)=(1.36; -0.92, -1.31)$  for
$\sigma_8 = 0.74$ and $(M_c; w_0, w_a)=(1.56; -0.81, -2.5)$ for
$\sigma_8 = 0.9$; and for $g^c(z_s)$ case, the best fit results
are $(M_c; w_0, w_a)=(1.38; -0.89, -1.21)$  for $\sigma_8 = 0.74$
and $(M_c; w_0, w_a)=(1.54; -0.83, -2.22)$ for $\sigma_8 = 0.9$;}
\end{figure}

\begin{figure}
 \epsscale{0.9} \plotone{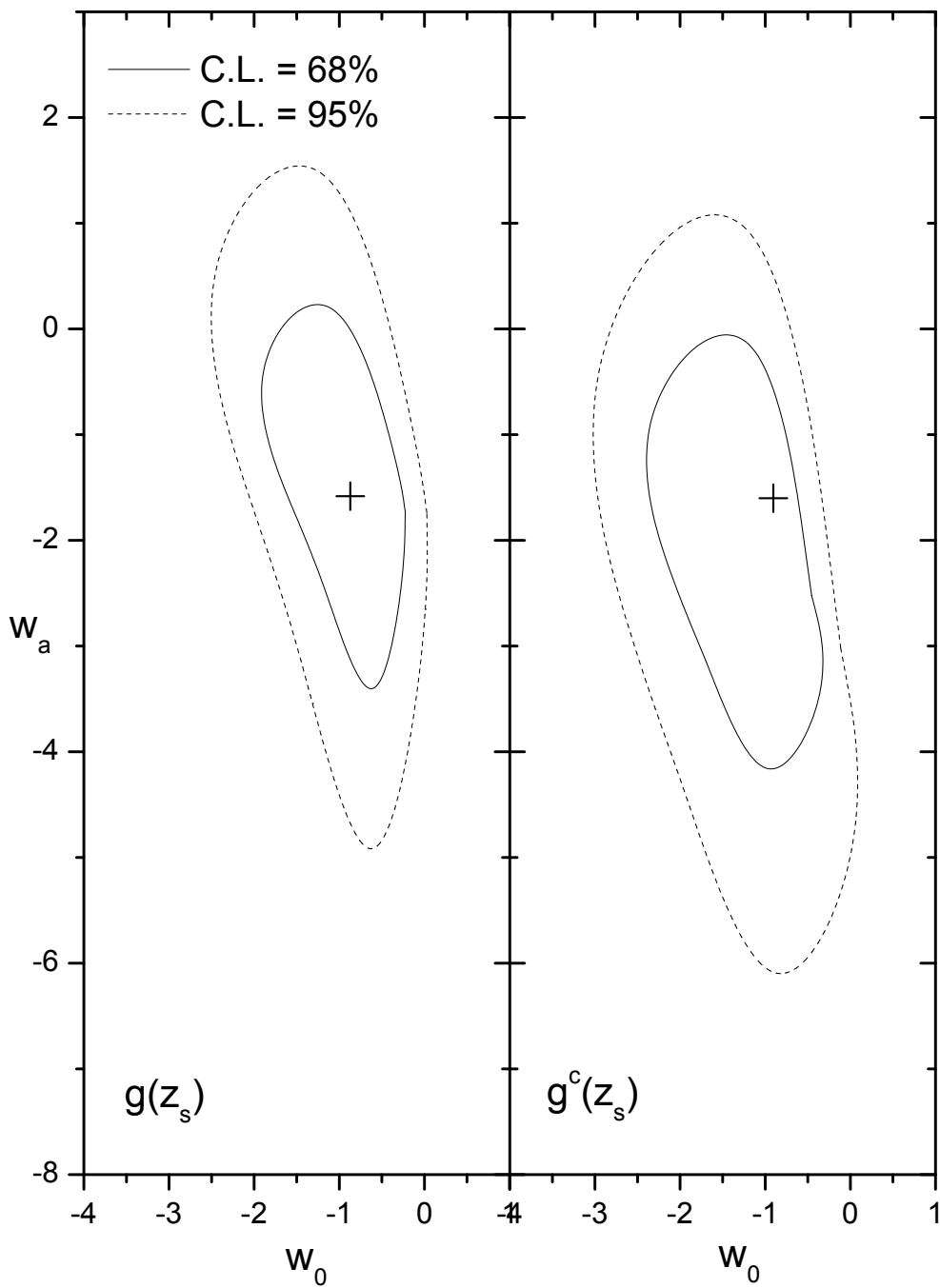}
\figcaption{\small 68\% C.L. and 95\% C.L.  allowed regions from the
CLASS statistical sample for the source redshift distribution
$f(z_s)$. The crosshirs mark the best-fit points  $(w_0,
w_a)=(-0.88,
 -1.55)$ for $g(z_s)$ case and $(w_0,
w_a)=(-0.91,
 -1.60)$ for $g^c(z_s)$ case .  The cosmological parameters
$(\Omega_M, h, \sigma_8)$  and the model parameter $M_c$ have been
marginalized}
\end{figure}

\end{document}